\documentstyle[aps,preprint]{revtex}

\begin{document}

\title{Fourier Path Integral Monte Carlo Method for the Calculation
of the Microcanonical Density of States}

\author{David L. Freeman}
\address{Department of Chemistry, University of Rhode Island, Kingston,
RI 02881}
\author{J.D. Doll}
\address{Department of Chemistry, Brown University, Providence, RI
02912}

\maketitle
\begin{abstract}
Using a Hubbard-Stratonovich transformation coupled with
Fourier path integral methods, expressions are derived for the numerical
evaluation of the microcanonical density of states for quantum particles
obeying Boltzmann statistics.  A numerical algorithm
is suggested to evaluate the quantum density of states
and illustrated on a one-dimensional
model system.
\end{abstract}
\pacs{PACS numbers: }
Over the past decade there has been considerable progress in the
development of path integral\cite{feynman} approaches to computational quantum
statistical mechanics\cite{berne,dfb}.
With few exceptions\cite{berne1} the application of path integral methods have
been
restricted to simulations in the canonical ensemble.  Canonical
simulations are amenable to path integral treatments because the
canonical density matrix is formally identical to the quantum propagator in
imaginary time.
The purpose of this note is to show that path integral methods can
provide an algorithmic basis for microcanonical simulations as well.

We let $\Omega (E) dE$ represent the number of energy states between $E$
and $E + dE$.  Throughout this paper we will
suppress the dependence of the density of states on the number of
particles in the system $N$ and the system volume $V$ for notational
convenience.
For $N$
indistinguishable particles obeying Boltzmann statistics, we begin
with the Fourier path integral expression for the canonical partition
function\cite{dfb}
\begin{equation}
Q(\beta ) =  \frac {1}{N!}
\left ( \frac {m}{2 \pi \beta \hbar ^2} \right )
^{3N/2} \left ( \prod _{k=1}^{\infty} \prod_{j=1}^{3N}
 \frac {1}{\sqrt{2 \pi \sigma_{k,j}^2}}
\right ) \int d {\bf r} d{\bf a} \exp [-\sum_{k,j} a_{k,j}^2/2 \sigma_{k,j}^2 -
\beta
<\Phi>_{{\bf a}}]. \label{eq:ddm}
\end{equation}
In Eq. (\ref{eq:ddm}) we have suppressed the dependence of the partition
function on $N$ and $V$, $\beta = 1/k_B T$ with $T$ the temperature and
$k_B$ the Boltzmann constant, {\bf r} is a collective $3N$-dimensional
vector representing the coordinates of all the particles in the system,
\begin{equation}
\sigma_{j,k}^2=\frac{2 \beta \hbar ^2}{m (k \pi )^2},
\end{equation}
$m$ is the mass of the constituent particles, $a_{k,j}$ is
Fourier coefficient $k$
for particle $j$, and for the potential energy $\Phi ({\bf r})$ we define
\begin{equation}
<\Phi >_{{\bf a}} = \int_0^1 du \Phi [{\bf r} (u)]
\end{equation}
as the average of the potential energy over a particular path.  As
discussed elsewhere\cite{dfb}, each path is parameterized by a Fourier sine
series.
In practice a finite number, $k_{max}$, of Fourier coefficients are
actually
included in Eq.(\ref{eq:ddm}).

The connection
between the canonical partition function and the microcanonical density
of states is given by
\begin{equation}
Q(\beta ) = \int dE \Omega (E) e^{-\beta E} \label{eq:laplace}
\end{equation}
which is just a Laplace transform relation. Equation ({\ref{eq:laplace}) can be
inverted using
the Mellin inversion integral
\begin{equation}
\Omega (E) = \frac {1}{2 \pi i} \int_{C} d\beta e^{\beta E} Q (\beta
) \label{eq:mellin}
\end{equation}
where the subscript $C$ on the integral in Eq. (\ref{eq:mellin}) implies
integration along a contour to the right and parallel to the imaginary
axis in the complex plane.
The inversion in the present application
is not simple, because of the inverse $\beta$-dependence in
the exponent of Eq. (\ref{eq:ddm}).  The $\beta$-dependence in
Eq.(\ref{eq:ddm}) can be altered using a
Hubbard-Stratonovich transformation~\cite{gold}.  Writing
\begin{equation}
\exp (-\sum_{k,j} a_{k,j}^2/2\sigma _{k,j}^2) = \prod_{k,j}
\left ( \frac {\sigma _{k,j}^2}{\pi}
 \right )^{1/2}\int d\mbox{\boldmath
$\phi$ \unboldmath} \exp (-\sum_{k,j}
\sigma_{k,j}^2 \phi_{k,j}^2 + i \sqrt{2} \sum_{k,j} a_{k,j} \phi_{k,j} )
\end{equation}
the expression for the partition function becomes
\[
Q(\beta) =
\frac {1}{N!} \left ( \frac {1} {\sqrt{2} \pi } \right ) ^{3N k_{max}}
\left ( \frac {m}{2 \pi  \hbar ^2} \right )
^{3N/2} \frac {1}{\beta^{3N/2}}
\]
\begin{equation}
\times \int d{\bf a} d {\bf r} d \mbox{\boldmath
$\phi$ \unboldmath}
 \exp [ -\beta (<\Phi>_{{\bf a}} + \sum_{k,j} \bar{\sigma}_{k,j}^2 \phi
_{k,j}^2) + i \sqrt{2} \sum_{k,j} a_{k,j} \phi_{k,j}
] \label{eq:qcan}
\end{equation}
where
\begin{equation}
\sigma _{k,j}^2= \beta \bar{\sigma}_{k,j}^2.
\end{equation}
At the expense of the introduction of the auxiliary variables $\mbox{\boldmath
$\phi$ \unboldmath}$, the $\beta$-dependence of the exponent in the
expression for the canonical partition function appears entirely in the
numerator. Using Eq.(\ref{eq:qcan}) the canonical partition function
can be introduced in the Mellin inversion integral, and the integration
with respect to $\beta$ can then be evaluated analytically~\cite{kubo}
\[
\Omega (E) = \frac {1}{\Gamma (3N/2)}
\frac {1}{N!} \left ( \frac {1} {\sqrt{2} \pi } \right ) ^
{3N k_{max}} \left ( \frac {m}{2 \pi  \hbar ^2} \right )
^{3N/2}
\]
\begin{equation}
\times \int d{\bf a} d {\bf r} d \mbox{\boldmath
$\phi$ \unboldmath} \cos ( \sqrt{2} \sum_{k,j} a_{k,j} \phi _{k,j} )
\Theta [E - <\Phi>_{{\bf a}} - \sum_{k,j} \bar{\sigma}_{k,j}^2 \phi
_{k,j}^2 ] [E - <\Phi>_{{\bf a}} - \sum_{k,j} \bar{\sigma}_{k,j}^2 \phi
_{k,j}^2 ]^{3N/2 -1 }, \label{eq:workm}
\end{equation}
where $\Gamma (x)$ is the gamma function and $\Theta (x)$ is the step
function.
Equation (\ref{eq:workm}) is the principal result of this note.
It is easy to verify that Eq.(\ref{eq:workm}) reduces to the proper free
particle and classical limits.

A possible approach to evaluate Eq.(\ref{eq:workm}) numerically is based
on the existence\cite{sw,whetten} of
methods to compute the
classical density of states. We then correct the
classical density of states by calculating the ratio of the quantum to
classical density.
If we specialize Eq.(\ref{eq:workm}) to a one-dimensional system and
examine the classical limit, it is easy to show that
the ratio of the
quantum to classical density of states can be written
\begin{equation}
\frac {\Omega (E)}{\Omega_{cl} (E)} =
\left ( \frac {1}{\sqrt{2} \pi} \right )^{k_{max}}
 \frac {<\cos
(\sqrt{2}\sum_ka_k\phi _k )>}{<\delta ({\bf a}) \delta (
\mbox{\boldmath $\phi$ \unboldmath})>} \label{eq:Rat}
\end{equation}
where the averages are taken with respect to the weight function
\begin{equation}
w(x,{\bf a},\mbox{\boldmath $\phi$ \unboldmath})=
\frac {\Theta [E -<\Phi>_{{\bf a}} -
\sum_k \bar{\sigma}_k^2 \phi_k^2]}{[E -<\Phi>_{{\bf a}} -
\sum_k \bar{\sigma}_k^2 \phi_k^2]^{1/2}}.
\end{equation}

In Figure 1  we present the
quantum density of states
for a Morse
oscillator
$
\Phi (x) = D_e \{1 - \exp [-a(x-x_e)] \}^2.
$
The calculations in Figure 1 were evaluated from Eq.(\ref{eq:Rat}) and
consisted of one hundred million Monte
Carlo points with $k_{max}=2$. The classical density of states was
determined from the classical limit of Eq.(\ref{eq:workm}) and was evaluated
using {\em Mathematica}~\cite{wolf}.
 For the potential parameters
we have chosen $a=1$, $x_e =1$, $m=1822.83$ au, and we have chosen
$D_e$ so that
the oscillator
frequency
$
\omega = ( 2 a^2 D_e/m  )^{1/2}
$
equals .006 au. The delta functions in Eq.(\ref{eq:Rat}) were
represented by gaussians with standard deviations of 0.5 au.
The path average of the potential energy for the Morse oscillator was
evaluated with trapezoid rule quadrature with 16 quadrature points.
The peaks in the density of states
seen in Figure 1 are in good
agreement with the analytic eigenvalues at $E/\hbar \omega = .488, 1.40$
and
2.21.
The finite widths of the peaks are a result of the gaussian
representation of the delta functions and the finite number of Fourier
coefficients used.
These preliminary results are encouraging, and the
investigation of several methods to evaluate Eq.(\ref{eq:workm}) for
more complex systems is in progress.

We would like to thank Professor Robert Topper for helpful discussions.
Acknowledgement is made to the Donors of the Petroleum Research Fund of
the American Chemical Society for support of this work.  Purchase of the
computer equipment used in this work was made possible by NSF grant
number CHE-9203498.

\newpage
\section*{Figure Captions}
\begin{enumerate}
\item
The quantum density of states in atomic units for
a one-dimensional
Morse oscillator with parameters defined in the text.
The error bars are at the double standard deviation level.
The maxima in the density of states are in good agreement with the
analytic eigenvalues at $E/\hbar \omega=$ .488,1.40 and 2.21.

\end{enumerate}
\newpage
\begin{figure}
\begin{center}
\setlength{\unitlength}{0.240900pt}
\ifx\plotpoint\undefined\newsavebox{\plotpoint}\fi
\sbox{\plotpoint}{\rule[-0.175pt]{0.350pt}{0.350pt}}%


\end{center}
\end{figure}

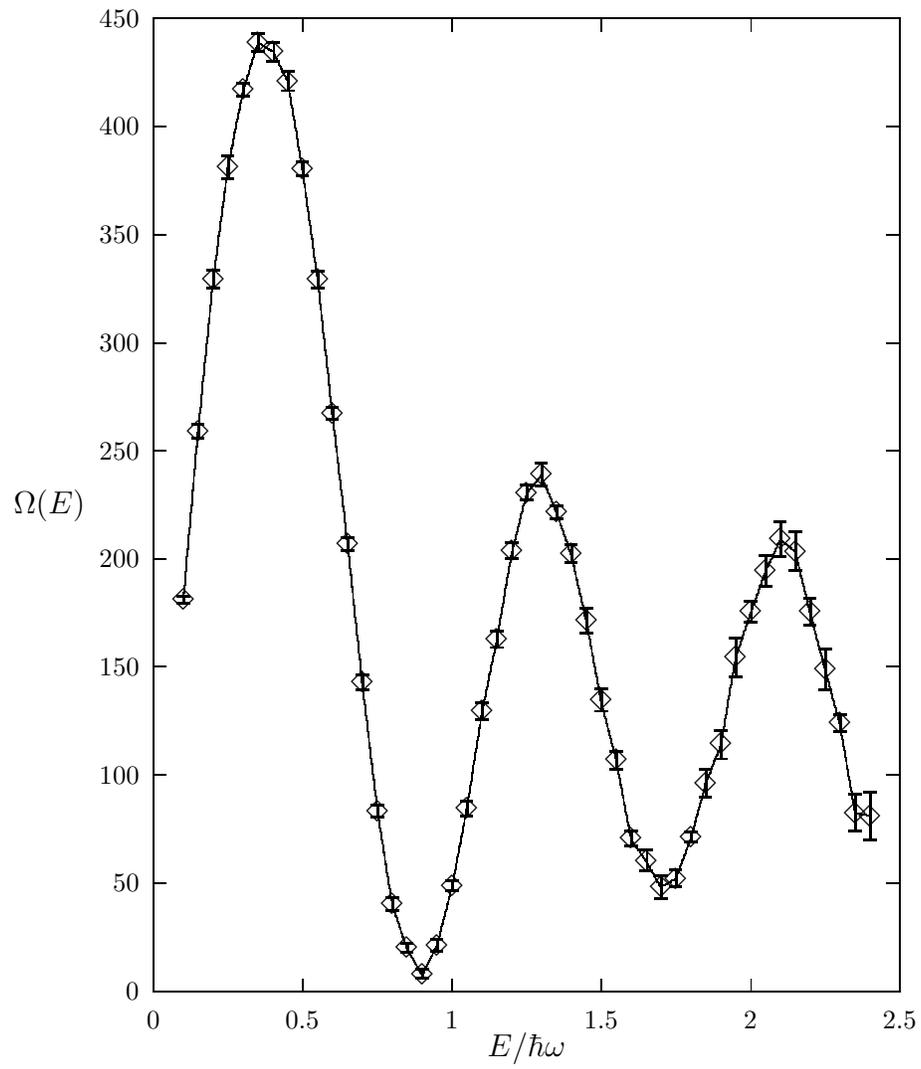
\end{document}